# mlf-core: a framework for deterministic machine learning

Lukas Heumos[1,2,3,4]*, Philipp Ehmele[5], Luis Kuhn Cuellar[1], Kevin Menden[1], Edmund Miller[6], Steffen Lemke[1], Gisela Gabernet[1#§], Sven Nahnsen[1,7,#§]

[1] Quantitative Biology Center (QBiC) Tübingen, University of Tübingen, Tübingen, Germany
[2] Institute of Computational Biology, Helmholtz Zentrum München, Munich, Germany
[3] Institute of Lung Biology and Disease and Comprehensive Pneumology Center, Helmholtz Zentrum München, Member of the German Center for Lung Research (DZL), Munich, Germany
[4] School of Life Sciences Weihenstephan, Technical University of Munich, Munich, Germany
[5] Department of Informatics, University of Hamburg, Hamburg
[6] Department of Biological Sciences and Center for Systems Biology, The University of Texas at Dallas, Richardson, Texas, USA
[7] Biomedical Data Science, Department for Computer Science, University of Tübingen, Tübingen, Germany
§ These authors contributed equally and share the senior authorship
* Corresponding author. E-mail: lukas.heumos@helmholtz-muenchen.de
# Corresponding author: E-mail: gisela.gabernet@qbic.uni-tuebingen.de
# Corresponding author: E-mail: sven.nahnsen@qbic.uni-tuebingen.de

## Abstract

Machine learning has shown extensive growth in recent years and is now routinely applied to sensitive areas[1]. To allow appropriate verification of predictive models before deployment, models must be deterministic. However, major machine learning libraries default to the usage of non-deterministic algorithms based on atomic operations. Solely fixing all random seeds is not sufficient for deterministic machine learning. To overcome this shortcoming, various machine learning libraries released deterministic counterparts to the non-deterministic algorithms. We evaluated the effect of these algorithms on determinism and runtime. Based on these results, we formulated a set of requirements for deterministic machine learning and developed a new software solution, the mlf-core ecosystem, which aids machine learning projects to meet and keep these requirements. We applied mlf-core to develop deterministic models in various biomedical fields including a single cell autoencoder with TensorFlow, a PyTorch-based U-Net model for liver-tumor segmentation in CT scans, and a liver cancer classifier based on gene expression profiles with XGBoost.

# Introduction

In recent years, machine learning has seen applications in almost all areas of the sciences and impacts society even in hidden ways, such as in the assessment of loan eligibility[2], clinical decision support[3], and crime or terrorist detection[4]. To adhere to democratic transparency standards and due to the sensitive application areas, reproducibility has been identified as a key factor to consider when applying machine learning[5,6]. Reproducibility in machine learning was defined by Gunderson and Kjensmo as "the ability of an independent research team to produce the same results using the same AI method based on the documentation made by the original team"[7].

Although the field of machine learning is progressing at a rapid pace, not all findings can be verified and reproduced[6,8]. Collberg and Proebsting in 2016 evaluated 402 computational experimental papers and could only reproduce 48.3% even when communicating with the authors[9,10]. Gunderson and Kjensmo report that for artificial intelligence papers 'only between a fifth and a third of the variables required for reproducibility are documented'[7]. The reasons for irreproducible machine learning are diverse, including unavailable (raw) data, unpublished code, unreported hyperparameters and missing reports on the hardware and software used for the analysis[8,11,12].

We argue that determinism should be equally considered in machine learning. Deterministic machine learning can be defined as the ability of an independent researcher to reproduce the *bit-exact* same results based on the complete documentation by the original team together with the model code when run on the same computational infrastructure.

Deterministic models have advantages beyond reproducibility which allow reproducing the exact same model metrics and predictions when the same training data is provided, and ease debugging by accurately describing metrics and predictions without the influence of unknown random factors. Moreover, they enable targeted experimentation, where the model performance cannot be attributed to random seeds. Scientists, developers and the general public benefit from deterministic machine learning, as a step towards building trustworthy machine learning[13].

Recently, several papers have compiled the requirements for reproducibility of machine learning algorithms and provided reproducibility checklists[14,15,16,5]. Even when adhering to these checklists, and reporting the utilized hardware and software together with the corresponding code, deterministic models are not necessarily obtained. This is especially relevant when the models are trained on graphical processing units (GPUs), which employ non-deterministic functions based on atomic operations[17]. Several machine learning frameworks exist that aim at solving the reproducibility issue. Projects such as Guild AI (https://guild.ai/) and Sacred (https://github.com/IDSIA/sacred) allow tracking all performed experiments and parameters used, and visualize those on a dashboard. Polyaxon (https://polyaxon.com/) and MLFlow (https://www.mlflow.org) additionally introduce containerization to ensure that the same software environment is used at runtime (Table 1). However, none of the machine learning experiment tracking frameworks currently track the hardware employed for the analysis or ensure that only deterministic functions are employed during training and inference. Therefore, there is a need for a single, user-friendly framework incorporating all requirements for deterministic machine learning.

| Framework | Tracking | Visualization | Container | Hardware | Determinism |
| --- | --- | --- | --- | --- | --- |
| Polyaxon | Full | Dashboard | Docker | No | No |
| Guild AI | Full | Dashboard | No | No | No |
| Sacred | Full with dependencies | Dashboard | No | Limited | No |
| MLflow | Full with models | Dashboard | Conda, Docker | No | No |

*Table 1. Comparison between different machine learning experiment tracking frameworks. Tracking includes capabilities of monitoring experiments, hyperparameters, metrics, artifacts, models or software dependencies. Visualization denotes the possibility of visually keeping track of experiments and automatic plotting abilities. Containers allow for consistent runtime environments. Hardware describes the competence of tracking the used hardware (model, quantity and generation) for any run. Sacred tracks the available hardware, but not solely the used hardware. Determinism describes the ability to statically verify that the code is deterministic.*

Here, we present mlf-core, a machine learning framework that enables building fully deterministic and therefore also reproducible machine learning projects. mlf-core is based on MLflow for machine learning experiment tracking, visualization and model deployment. Additionally, mlf-core provides project templates and static code analysis (linting) functionality that ensures the sole usage of deterministic algorithms for GPU computing as well as setting all necessary random seeds for deterministic results. We identified the necessary settings to ensure determinism by analyzing the hardware- and software-derived sources of non-determinism in three machine learning models built with the PyTorch[18], TensorFlow[19] and XGBoost[20] libraries. We concluded that setting random seeds alone was not sufficient to ensure determinism in GPU-based computations. When enabling the full deterministic requirements, deterministic results were obtained for all models.

We showcase the applicability of mlf-core to three use cases for various biomedical applications. First, using an autoencoder model built with TensorFlow, we demonstrate how non-deterministic model training and inference can strongly influence the outcome of algorithms commonly used to analyze single-cell RNA-sequencing (scRNA-seq) data. Second, we exhibit how feature importance varies for non-deterministic runs with a liver cancer classifier based on transcriptomics data implemented with XGBoost. Third, we implemented a U-Net[21] model with Pytorch for liver-tumor segmentation in abdominal CT scans showing that only deterministic training yields the exact same model predictions.

Together with providing access to FAIR[9] (Findable, Accessible, Interoperable, Reusable) training data and code, mlf-core enables researchers to perform deterministic machine learning research.

## Results

*The mlf-core ecosystem for deterministic machine learning*

To address the need for an ecosystem that enables the easy setup of deterministic machine learning models, we developed mlf-core. mlf-core is inspired by nf-core[22], a framework for building reproducible bioinformatics pipelines written in Nextflow; and cookietemple[23], a collection of best-practice templates for various domains and programming languages. mlf-core provides a complete solution for deterministic machine learning that ensures deterministic model training, and tackles other important aspects for determinism of the developed models, such as result visualization and hyperparameter tracking, operating system tracking, dependency management with containers, and a publicly-hosted model documentation (Figure 1).

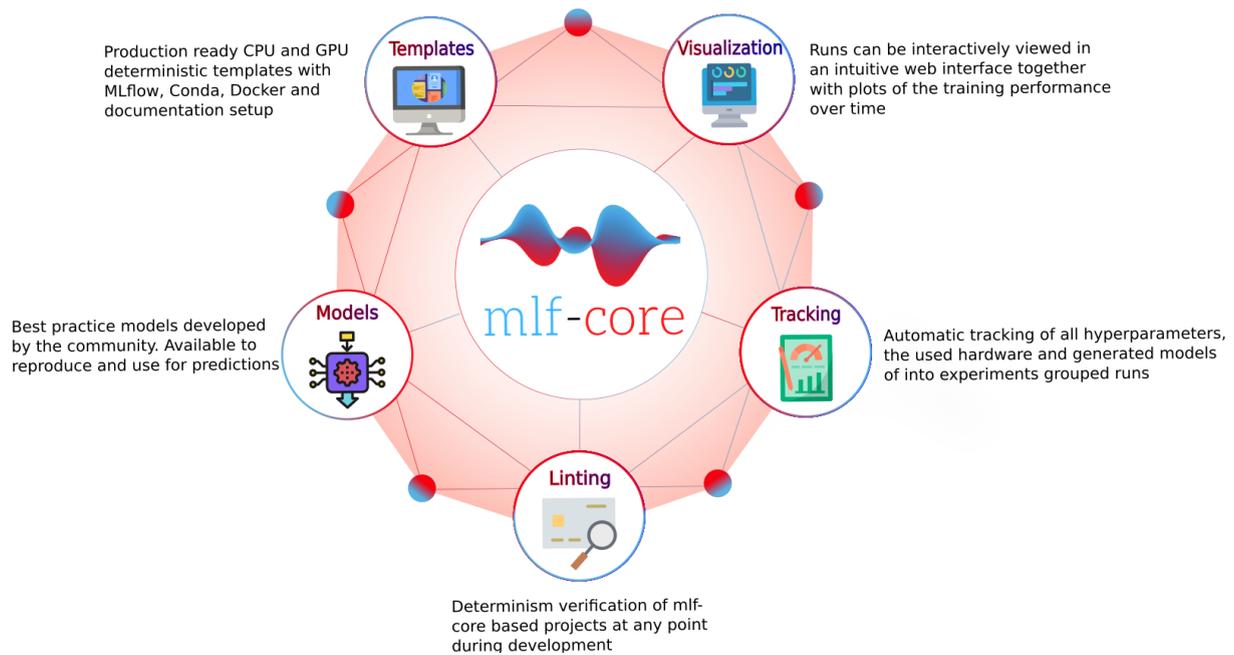

Figure 1. Overview of the mlf-core ecosystem. The mlf-core ecosystem comprises the Python packages mlf-core and system-intelligence, community developed reproducible mlf-core models and a set of GPU enabled Docker containers.

mlf-core allows for the interactive creation of deterministic machine learning projects based on best-practice templates for the most widely used machine learning frameworks, namely PyTorch, TensorFlow or XGBoost. These templates include a minimal running example for training e.g. a convolutional neural network or a distributed gradient boosting model which meet a set of requirements that have been determined and empirically verified to enable deterministic machine learning model training on both CPUs or GPUs. These requirements include: The containerization of the runtime environment, the setting of all random seeds for all used libraries, the usage of solely deterministic algorithms and the tracking of all hyperparameters and metrics.

A custom code linter is provided as part of the mlf-core Python package that checks whether already developed projects adhere to the determined requirements for deterministic machine learning. If any setting in the code potentially leads to non-deterministic models (e.g. not all required random seeds are set, deterministic algorithms are not enforced or known non-deterministic algorithms are used), the user is warned and possible deterministic alternatives are proposed (Figure 2). Enabling and disabling determinism modes allows researchers to quantify the impact of non-deterministic training and prediction in downstream analysis, such as statistical significance evaluations.

Furthermore, the hardware used for training is automatically logged and reported with system-intelligence, a tool of the mlf-core ecosystem, even in distributed environments with different hardware architectures. All mlf-core project templates include an MLflow setup to use Conda[24] or Docker[25] containers for dependency management, which are recommended to ensure interoperability in a variety of infrastructures. Continuous integration workflows based on GitHub Actions (https://github.com/features/actions) enable automatic building of the Docker container, and statically verify code quality (linting). Sphinx (https://www.sphinx-doc.org) is employed for documentation building in order to encourage developers to document their machine learning project. The mlf-core project creation process ends with the option to automatically upload the model code to a newly created GitHub repository to encourage code sharing (Figure 2). Existing projects are kept up to date with the mlf-core sync functionality (see Methods) allowing for regular updates to ensure determinism and adherence to current best-practices. Already existing projects are easily transformed into mlf-core projects by solely copying the model code into the corresponding section in the template. To enable easy model scores and hyperparameter tracking, mlf-core employs MLflow to provide an overview of all the training runs and automatically log all hyperparameters and obtained metrics for each run as well as an instance of the trained model. The automatically saved Tensorboard events can be used to visualize the model graph and tensor trajectories (Supplementary Figure 1).

mlf-core's tight integration of MLflow also allows for models to be easily served as representational state transfer (REST) APIs, effectively bridging research and application by allowing users to request predictions of the model. Model training and inference can be performed in various infrastructure including all common Cloud providers (e.g. Amazon Web Services, Google Cloud) and scales to big data with Apache Spark[26].

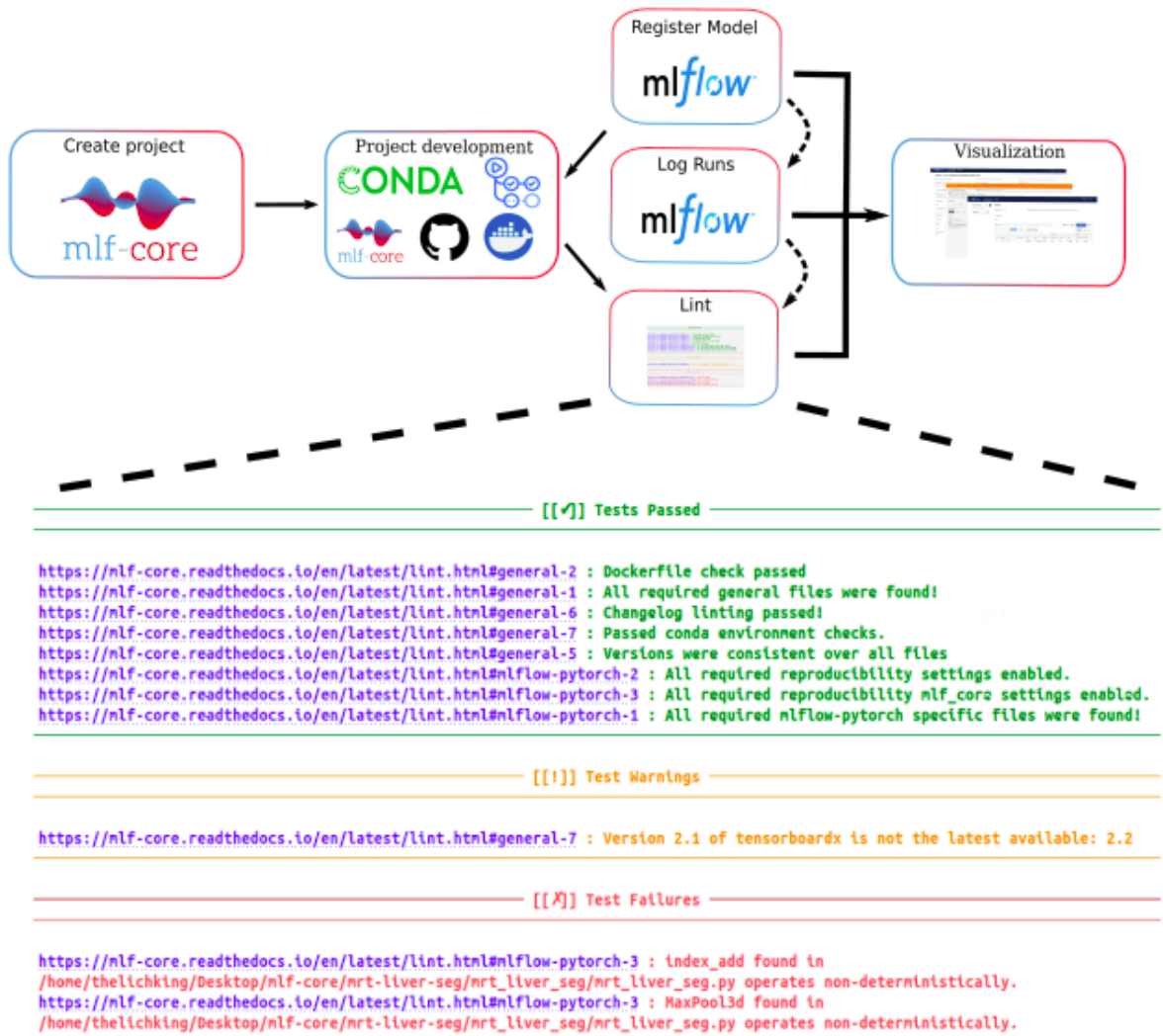

*Figure 2. Workflow for the development of a mlf-core based model. Projects are created interactively with the mlf-core command line tool, and automatically published to a newly created GitHub repository. Project development is facilitated with the use of a continuous integration pipeline using GitHub Actions and isolated runtime environments based on Conda and Docker. mlf-core lint statically validates determinism and informs on violations. All training runs log all metrics and save models in a specified model registry. The MLflow web interface and Tensorboard facilitate interactive exploration of the generated models. A tutorial for the complete workflow is available on https://mlf-core.readthedocs.io/en/latest/tutorial.html .*

*Validation of machine learning determinism with mlf-core*

To empirically validate that the mlf-core ecosystem ensures the implementation of deterministic models, we assessed the determinism of three machine learning models implemented with and without the determinism requirements employed by the mlf-core ecosystem. A 2-layer convolutional neural network (CNN) model was trained to classify digits using the MNIST dataset, programmed with the PyTorch[18] and TensorFlow[19] libraries. Determinism for gradient boosted trees implemented with XGBoost[20], a distributed gradient boosting library, was evaluated using the Covertype dataset. All evaluations were conducted on systems using a CPU, a single GPU and multiple GPUs (Supplementary Table 1). Additionally, three different training setups were tested: (1) A setup without specified seeds or any deterministic setting (*random*), (2) a setup that specified random seeds for all libraries used for the model implementation (*seeds*), and (3) a setup which specified all random seeds, and additionally forced deterministic algorithms together with disabled cuDNN benchmark functionality, as implemented in the mlf-core ecosystem (*deterministic*) (Supplementary Figure 2). Figure 3a shows the loss value of the last training iteration of the TensorFlow 2-layer CNN implementation for five training repetitions. Only the deterministic setup implemented with mlf-core achieved fully deterministic results on all tested infrastructures, including a single CPU, a single GPU and a multi-GPU setup (Figure 3a for the TensorFlow implementation, Supplementary Figures 3 and 4 for the PyTorch and XGBoost implementations, respectively and Supplementary Figure 11 for the weight standard deviations).

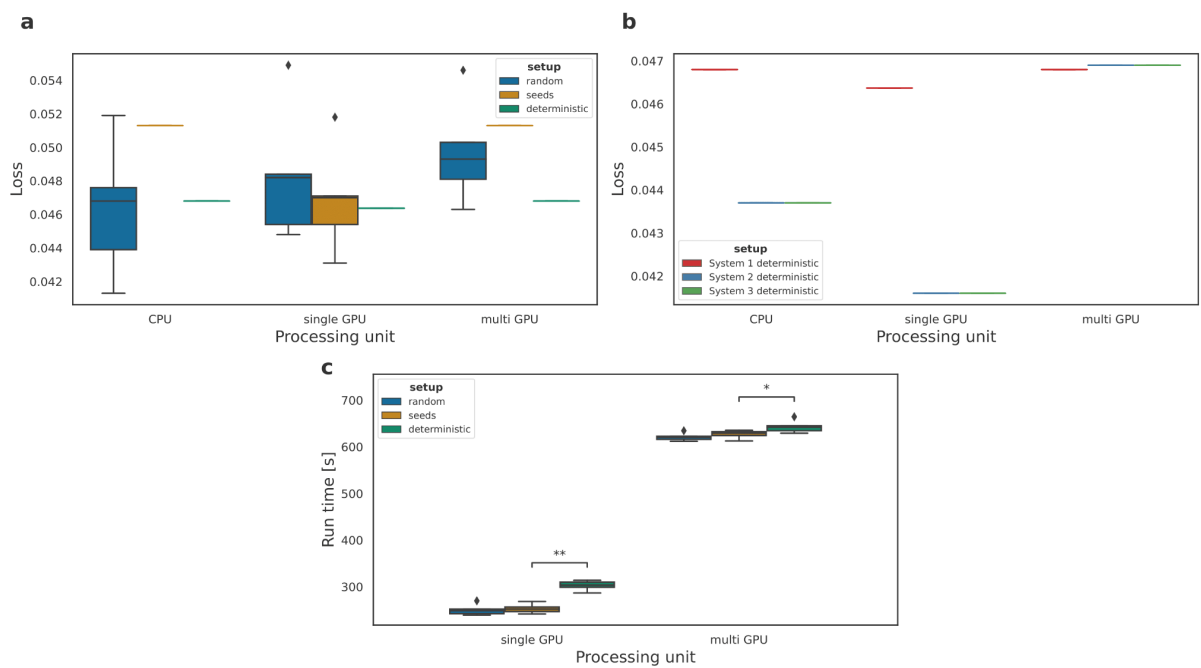

*Figure 3. Determinism evaluation of a convolutional neural network with dropout layers implemented in TensorFlow 2.2 and trained on the MNIST dataset. (a) Loss variation across five training runs (N=5) after 25 training epochs. Different settings were compared: without random seeds or deterministic algorithms (random), solely setting the same library random seeds across runs (seeds); or setting the random seeds and enabling the deterministic algorithms for TensorFlow as implemented in the mlf-core ecosystem (deterministic). Training runs on CPU, single GPUs and multi-GPU setup in the same system are compared. (b) Loss comparison of five runs (N=5) across individual systems with different hardware (systems 1 and 2/3), and individual systems with the same hardware (systems 2 and 3), all with deterministic settings. (c) Training run time for 25 epochs when training the model without setting random seeds, when setting the random seeds and when forcing deterministic algorithms. For each setting, 5 individual runs were considered (N=5).*

Even when all deterministic settings were enabled, the same models trained on systems with different hardware generated different results, emphasizing the need of hardware tracking when training models (Figure 3b). Training the model in two machines with identical hardware resulted in the exact same loss as demonstrated with systems 2 and 3 (Figure 3b). These deterministic results could be reproduced for PyTorch (Supplementary Figure 3). The evaluation of the XGBoost library unveiled non-determinism when using XGBoost version 1.0.2 and XGBoost versions 1.1.0 compiled with CUDA 9 (Supplementary Figure 4). However, training models with XGBoost version 1.1.0 compiled with CUDA 10 on a single GPU led to fully deterministic results. Regardless, runs with multiple GPUs on all systems with XGBoost did not result in deterministic runs, suggesting that XGBoost models should be trained on single GPUs to ensure determinism (Supplementary Figure 4).

Next, we investigated the influence of forced determinism requirements on model training runtimes. Enabling *deterministic* algorithms resulted in significantly higher run times both when training on a single GPU (two-tailed t-test, *p-value=0.001*) and multiple GPUs (two-tailed t-test, *p-value=0.014*) when comparing the *seed* and *deterministic* settings for TensorFlow (Figure 3c). Runs with PyTorch and deterministic algorithms enabled did not result in significantly higher runtimes compared to the *seeds-only* setting (two-tailed t-test $P=.5558$) (Supplementary Figure 3) for a single GPU. For multiple GPUs, the runtime was significantly higher when *deterministic* settings were enabled (two-tailed t-test $P=0.001$). These results indicate that deterministic models might result in longer runtimes.

*mlf-core biomedical use cases*

To showcase the applicability and advantages of the mlf-core ecosystem for deterministic and reproducible machine learning in various biomedical fields, we implemented three use cases, each one based on one of the three machine learning libraries supported by mlf-core.

*mlf-core enables reproducible analysis of scRNA-seq data with TensorFlow*

Deep learning-based methods are commonly used for the analysis of scRNA-seq data, which is ideally suited for machine learning due to high sample numbers ranging from thousands to millions of cells. For instance, unsupervised methods such as autoencoders have been successfully applied in a variety of settings[27–30]. Given the frequent use of deep learning models for scRNA-seq data, we wanted to evaluate how randomness affects the reproducibility of downstream analyses. To test this, we fitted a simple autoencoder model to a peripheral blood mononuclear cell (PBMC) scRNA-seq dataset. The comparison of the loss after and during training shows differences for non-deterministic experiments, even if no convolutional layers are used (Figure 4a, Supplementary Figure 5a, b and d). We furthermore used the autoencoder latent space as input to Uniform Manifold Approximation and Projection (UMAP)[31], a frequently used algorithm for visualizing scRNA-seq data in two dimensions. Non-deterministic models lead to visibly different UMAP plots, which hinders comparison and thus reproducibility (Figure 4b). Finally, we performed clustering of cells based on the latent space embeddings, which is essential to most scRNA-seq data analyses. Comparison of cluster sizes for different runs showed considerable differences among the non-deterministic runs which would result in biologically relevant different cell type annotations (Figure 4c). When applying the deterministic settings, as provided by the mlf-core framework, the autoencoder loss reached always the same value among the different runs (Figure 4a). Moreover, the cluster sizes and UMAP projection remained identical (Figure 4b and c).

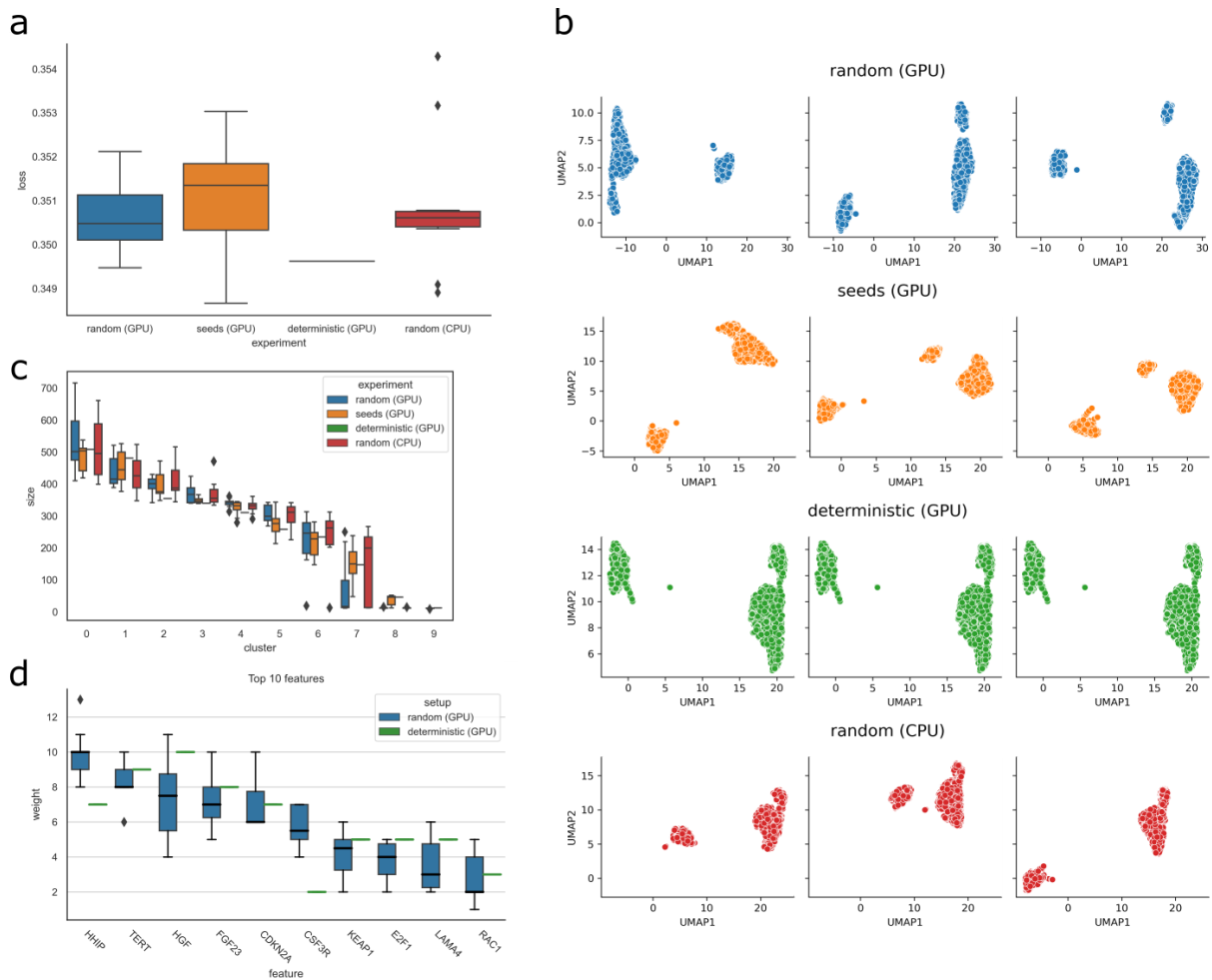

Heumos et al.*Figure 4. Impact of non-deterministic settings on an autoencoder model for single-cell RNA-seq data and on a XGboost classification model on a liver cancer dataset (a) Loss variation of an autoencoder trained on single-cell RNA-seq data after 1000 epochs of training with no random seeds set (random), all possible random seeds set (seeds) and deterministic algorithms forced (deterministic). For each setting, ten different runs were performed (N=10). (b) UMAP plots generated from the autoencoder embedding after training for 1000 epochs in different experimental settings. Each row corresponds to a different setting and each column to a different run with the same settings. (c) Boxplots of cluster sizes for different experimental settings. Clusters were generated using the Leiden-clustering algorithm and the autoencoder embedding. Different clusters are marked at the x-axis, boxplot colors correspond to experimental settings. (d) The ten features with the highest average weight of the non-deterministic XGBoost model on liver cancer. For each setup ten runs were performed (N=10). The features were ordered by decreasing average weight.*

*mlf-core enables deterministic feature importance determination with XGBoost*

Machine learning can be applied to identify novel oncology projects where data on large patient cohorts is available to identify novel biomarkers for diagnostics or potential targets[32,33]. One widely used unsupervised approach to identify novel biomarkers is to train a model on gene expression data and to analyze the model feature importance assigned to the genes[34,35]. However, if no measures are taken to ensure reproducibility during the machine learning analysis, the importance assigned to the gene features will vary among different runs of the same model. To showcase this we applied a feature importance determination approach to hepatocellular carcinoma (HCC), the most frequent type of liver cancer[36]. Using an RNA-Seq derived gene expression dataset consisting of publicly accessible data, we trained a gradient boosting model (XGBoost) to classify gene expression profiles into malignant and healthy.

To evaluate the determinism of this model, we compared the loss values (log loss) among different runs with deterministic settings as provided by mlf-core, to the loss obtained from runs without deterministic settings (Supplementary Figure 5c). When determinism was not controlled, the model training loss values differed across the runs (Supplementary Figure 5c). The average weights of the ten most important features of non-deterministic XGBoost models show considerable variances across the runs (Figure 4d). When setting the seeds to obtain deterministic results, the loss values and feature important weights stayed constant across all runs (Figure 4d).

*mlf-core enables reproducible semantic segmentation of liver CT scans with PyTorch*

Computed tomography (CT) is commonly used for liver tumor evaluation and staging. It allows the study of anomalies in the shape and texture of liver tissue, which are important biomarkers for initial disease diagnosis and progression[37]. In this setting, volumetric segmentation of liver and malignant tissue in CT scans is an important task for cancer diagnosis and treatment, e.g. it allows the calculation of tumor burden (liver/tumor ratio) to assess disease progression and effectiveness of treatment[38].

We evaluated the performance and reproducibility of a U-Net architecture[39] trained on the *Liver Tumor Segmentation Benchmark* (LiTS) dataset[40], since this model has been widely applied for semantic segmentation of microscopy and medical imaging data[41–44]. The training dataset of LiTS comprises 131 abdominal CT scans of patients with HCC, with the corresponding ground-truth segmentation of liver and tumor lesions, as annotated by experts.

We randomly sampled 10% of the dataset to define a small test set (13 tomograms), and trained our models for 1000 epochs with the remaining tomograms using different experimental GPU setups. We compared the final loss between runs and did not observe any variation of the deterministic setup, as opposed to the *random* and *seed* setups (Figure 5a). We tested reproducibility of prediction by evaluating the performance of the models on the abovementioned test set, using *Intersection over Union* (IoU) as a metric (Figure 5b). Likewise, we observed no differences in IoU for the deterministic setup, while other setups exhibited noticeable variation, in particular for the tumor class. We calculated the standard deviation of the voxel-wise, softmax value used for prediction on the test set (Figure 5c), and could verify that these values are reproducible using the deterministic setup. Moreover, we used the predicted segmentation masks to compute tumor burden for each tomogram in the test set, and measured the standard deviation across runs (Figure 5d). We could only observe reproducible tumor burden calculations using the deterministic setup.

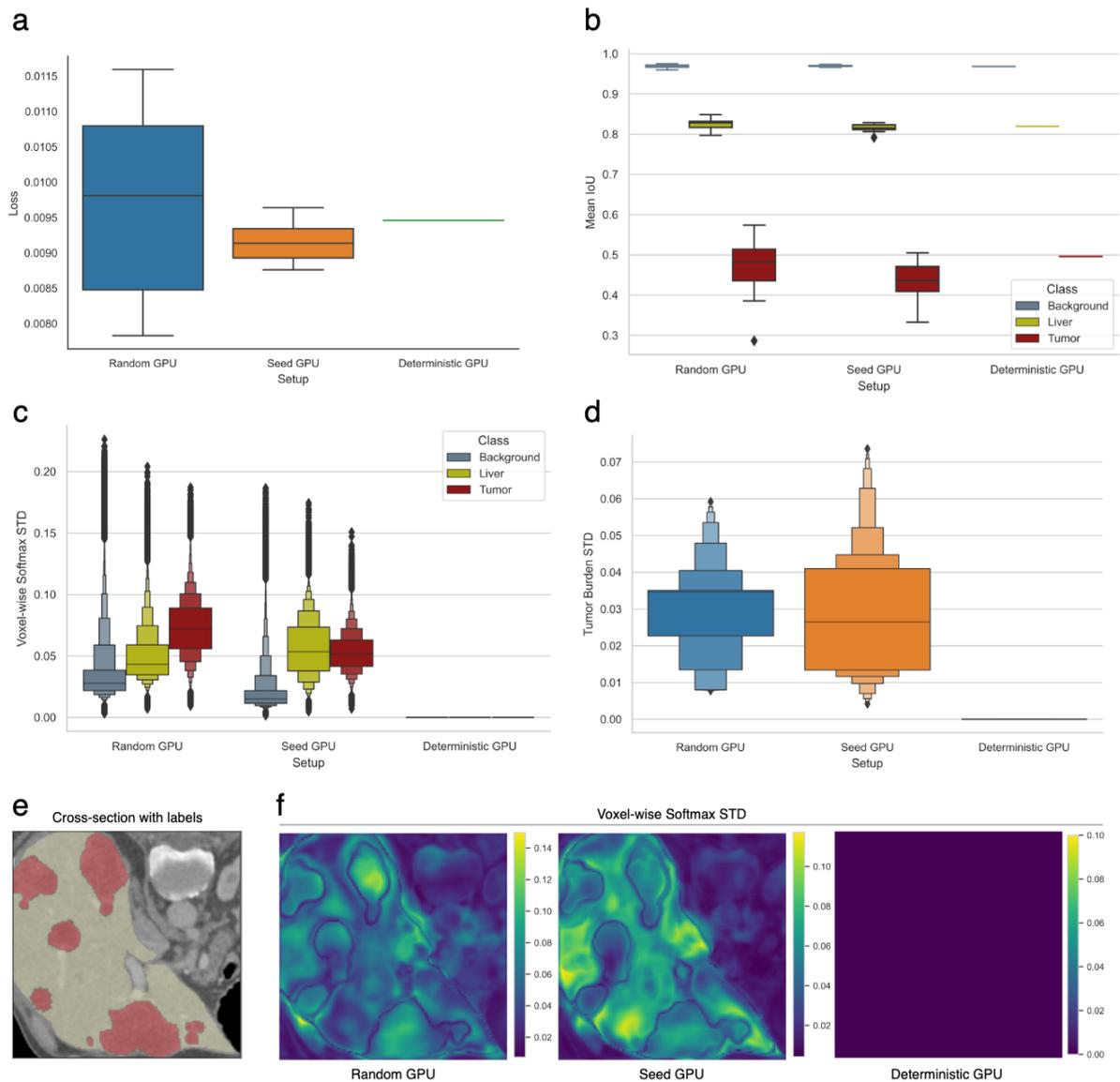

*Figure 5. Impact of non-deterministic settings on a U-Net model for semantic segmentation trained on the LiTS dataset and implemented with PyTorch. (a) Boxplot of final losses after 1000 epochs of training with no random seeds set (random), all possible random seeds set (seeds) and deterministic algorithms forced (deterministic). For each setting, ten different runs were performed (N=10). (b) Performance of the trained models on the testing set (13 tomograms) for different experimental setups, showing mean IoU values per class (N=10). (c) Letter-value[45] plot of standard deviations (STD) from the voxel-wise, softmax values used for prediction, as computed by trained models (Supplementary Figure 9) with different experimental setups (N=10). (d) Letter-value plot of standard deviation values of tumor burden measurements from the testing set, using the segmentation masks predicted by the trained models (N=10). (e) Cross-section of a tomogram from the test set with the ground truth class labels, color coding voxels for the liver class (yellow) and tumor class (red). (f) The same cross-section as (e), showing the standard deviation of the voxel-wise softmax values among the runs (N=10), for the three experimental setups.*

## Discussion

In the current work, we introduce a framework to ensure deterministic settings in machine learning projects. Setting random seeds for all the libraries during model training has proven necessary but not sufficient to ensure the bit-exact reproducibility of the model metrics and subsequently the performed predictions (Figure 3, Supplementary Figure 3 and 4). This is due to the fact that machine learning libraries use non-deterministic algorithms based on atomic operations when training with GPUs, to ensure the fastest training. The large number of available cores of GPUs allow for massively parallel processing of simple tasks such as additions or multiplications which are employed during the training of (convolutional) neural networks. However, the employed atomic operations result in non-determinism due to the non-associativity of floating-point arithmetics[46,47]. Applied in large numbers, *atomic add* based sum reduce algorithms lead to a significant accumulation of floating-point errors. Functions such as bias additions and max-pooling, which are based on atomic add and batch normalization, will therefore produce non-deterministic results[48]. When replacing the atomic add operations with deterministic equivalents the bias additions become deterministic, but potentially at a speed loss. Machine learning frameworks such as PyTorch, TensorFlow and XGBoost still allow for the usage of several less commonly used algorithms based on *atomic add* (https://pytorch.org/docs/stable/notes/randomness.html and https://github.com/NVIDIA/framework-determinism). So far, equivalent, deterministic algorithms are not available for these functions. Enforcing deterministic algorithms through machine library specific flags is also not sufficient, because many common algorithms (e.g. 3D convolution) are still implemented non-deterministically based on atomic operations. The mlf-core linting tool integrated into the mlf-core framework warns developers when non-deterministic algorithms are being used.

Caution should be taken when using functionality such as *cuDNN benchmark,* a function from the NVIDIA cuDNN library, which selects the most appropriate algorithm taking as a criteria the fastest implementation. *cuDNN benchmark* can choose among deterministic and non-deterministic algorithms, making it difficult to evaluate determinism as it can be switched on and off in different experiments. The observed determinism of TensorFlow (Figure 3a) when training with multiple GPUs may be due to *cuDNN benchmark* selecting deterministic variants for the used algorithms. However, it cannot be guaranteed that the same algorithms will be selected when training the models on different systems with the same hardware. Hence, deactivating *cuDNN benchmark* and enforcing deterministic algorithms is key to ensure the bit-exact same results. Since not all algorithms currently have deterministic equivalents, further work to implement deterministic versions of non-deterministic counterparts is imperative.

The selection of deterministic algorithms for all computations might come at the cost of a higher runtime necessary for training, as observed for the TensorFlow model trained on the MNIST dataset (Figure 3c)[48]. However, we did not observe increased runtimes while evaluating a similar model implemented in PyTorch, where the runtime was even lower with deterministic algorithms enabled (Supplementary Figure 3d). Moreover, runtimes of the significantly larger 3D U-Net model, trained on the LiTS dataset, and implemented in PyTorch, did not increase but were shorter in average (Supplementary Figure 10c). In fact, we argue that forcing deterministic algorithms even for extremely large-scale models will benefit development time in the long run since debugging, experimentation and regression testing are greatly simplified.

Data distribution over multiple worker instances with Dask[49], a cluster often employed to distribute over several GPUs, is not deterministic (Supplementary Figure 4). When training a model on multiple GPUs on a Dask cluster every GPU is treated as a standalone worker. Dask requires the data to be distributed and partitioned across workers, which for maximum speed currently does not guarantee determinism. Therefore, the usage of multiple worker instances either in the form of CPUs or GPUs is not recommended when deterministic machine learning is desired.

The software stack has to stay consistent to ensure determinism, since different CUDA drivers or machine learning algorithm implementations affect the generated results (Supplementary Figure 4). A further issue usually hidden to the end user, is that the distributed packages have to be compiled with the correct dependency versions. The hardware stack also plays a major role in determinism. Different GPU architectures partition the computational load for massively parallel processing in distinct ways. Any changes in this partitioning will be reflected in differences in floating-point rounding error accumulation[50,17] (Figure 3). Cloud computing greatly simplifies access to standardized hardware which together with mlf-core's logging of used hardware mitigate this issue allowing for portable deterministic machine learning.

We summarize the technical requirements to ensure deterministic results based on the aforementioned causes for non-determinism identified in this work as the following points:

- **Setting random seeds**: All random seeds used by all imported libraries need to be set.
- **Using exclusively deterministic algorithms:** The use of deterministic algorithms must be enforced and ensured that non-deterministic algorithms cannot be selected. Furthermore, functionality such as *cuDNN Benchmark*, which chooses algorithms at runtime based on their performance in different hardware, must be switched off.
- **Containerising the runtime environment:** The complete runtime environment must be containerized to ensure a consistent runtime environment and control the versions of all software used for model training.
- **Logging the hardware**: The used hardware architecture must be documented, especially the used CPU and all GPUs, and kept constant for reproducibility.
- **Documenting model training and parameters, and sharing the code:** All hyperparameters, obtained metrics and model details must be documented together with the full open source code for all model training runs.

The mlf-core framework enforces all the aforementioned requirements for deterministic machine learning directly into the model code, and thus provides the first complete solution for CPU and GPU reproducible machine learning (Supplementary Figure 7). The machine learning templates provided by mlf-core automatically set all the necessary random seeds for the Tensorflow, PyTorch or XGBoost libraries. The mlf-core linting functionality ensures that non-deterministic functions are not employed, and warns the user if selected functions are non-deterministic. mlf-core supports Docker to ensure a consistent runtime environment, as well as tracking of the hardware employed for ML training with the system-intelligence tool. The seamless integration with MLFlow ensures the documentation of hyperparameters, training metrics and model details for each of the training runs. The very high degree of automation and ease-of-use opens the framework to a wide audience.

We showcased the usage of the mlf-core ecosystem to develop deterministic machine learning models for three different use cases relevant for biomedical data science. Using TensorFlow and simple autoencoder models to process scRNA-seq, our data showed that non-deterministic algorithms have an impact on the reproducibility of scRNA-seq data analysis, leading to differences in cell clustering and thus potentially cell type-assignments.

The liver cancer classifier based on the mlf-core XGBoost template highlights the importance of deterministic algorithms and set seeds when extracting feature importance. Life scientists base further experiments on candidate gene importance and therefore need reproducible computational models especially when new data is regularly added to the experiment.

For semantic segmentation of CT scans using PyTorch, we showed that the deterministic setup is able to produce reproducible results using a simplified U-Net model, both during training and for prediction. Enforcing determinism of models used for liver-tumor segmentation of abdominal CT scans is critical to ensure reproducibility in patient diagnostics, especially when measuring tumor burden or applying the *Response Evaluation Criteria in Solid Tumor* (RECIST) protocol for tumor staging[51].

Our work unveils the causes of non-deterministic machine learning and we provide the first solution to enable deterministic predictions: the mlf-core framework. We envision mlf-core to enable the machine learning community to develop deterministic machine learning models, advancing research while increasing society's trust in deployed models.

# Methods

*Mlf-core ecosystem implementation*

The mlf-core ecosystem comprises the two Python packages mlf-core (https://github.com/mlf-core/mlf-core) and system-intelligence (https://github.com/mlf-core/system-intelligence), a collection of GPU enabled Docker containers (https://github.com/mlf-core/containers) and the corresponding website ([https://mlf-core.com](https://mlf-core.com)).

The mlf-core and system-intelligence Python packages are based on cookietemple's cli-python template[23]. The creation of the machine learning templates is primarily facilitated by cookiecutter (https://github.com/cookiecutter/cookiecutter) to replace variable defaults with user defined choices. The templates themselves are based on the determinism experiments conducted in this work, extended with MLflow (https://mlflow.org) integration.

All mlf-core templates are versioned allowing for the comparison of existing project's former template versions and the latest mlf-core release's template version. If newer versions are available a sync pull request incorporating all new changes is created against the existing project. This process uses the user's GitHub account together with the user's GitHub Token which was securely saved as a GitHub secret during the project creation step. In the same step a *TEMPLATE* branch is created, which at any time only contains the template code. When a template sync occurs the project template is recreated on the *TEMPLATE* branch and a pull request based on the new template is submitted against the development branch. A GitHub Actions based workflow runs on a daily basis to check for new template versions.

The verification of best-practices and machine learning determinism is facilitated through a custom static code analyzer, which checks the complete code against predefined sets of deterministic functions and points out non-deterministic functions when they are used. These sets are continuously updated and new developments lead to new mlf-core releases. The static code analyzer runs on every push and pull request event against any branch on GitHub or can be called manually with the *mlf-core lint* command.

The base Docker container for all mlf-core template Docker container is *nvidia/cuda:11.2.1-cudnn8-devel-ubuntu20.04* extended with Miniconda 4.9.2. Additionally, the PyTorch template defines the CUBLAS_WORKSPACE_CONFIG=:4096:8 environment variable to disable non-deterministic behavior of multi-stream execution in internal workspace selection for routines running in parallel streams.

*Evaluating machine learning library determinism*

All deterministic machine learning experiments were conducted with Nextflow version 20.01.0 build 5264 on the hardware systems shown in Table 2. Every experiment was conducted with Docker containers with a custom base inheriting from *nvidia/cuda:10.2-base-ubuntu18.04*. The full setup including Docker containers, Conda environments and code are available on GitHub: https://github.com/mlf-core/machine_learning_determinism_evaluation. The execution time of all trained models was calculated with Python's inbuilt *timeit* function.

PyTorch determinism evaluation setup

To test the deterministic settings of the PyTorch library, a neural network with 2 convolutional (32 and 64 neurons), 2 fully connected (128 and 10 neurons) and 2 dropout (0.25) layers was used. The respective activation functions were rectified linear functions followed by 2-dimensional maxpooling after the second convolutional layer. Optimization was conducted using the Adam optimizer, used to minimize a log-likelihood loss. Hyperparameters were left at their default values. The models were trained for 25 epochs with default weight initialization. The models were trained to classify images of the MNIST[52] dataset, which contains 70000 (60000 training and 10000 testing) handwritten 28 times 28 greyscale digits, into the correct digit.

All PyTorch experiments were conducted with PyTorch version 1.7.1 and multi GPU support was enabled using the *DataParallel* API.

To evaluate determinism on the PyTorch library, three different experimental setups were employed:

- *Random*: no random seeds were set.
- Seeds: Random seeds were set for NumPy, PyTorch, Python's Random module and the Python hash seed.
- Deterministic: All random seeds of the setup *seeds* were set together with the enforcing of all deterministic algorithms and cuDNN benchmark was disabled (Supplementary Figure 8).

Further, all functions labeled as knowingly non-deterministic by PyTorch (https://pytorch.org/docs/stable/notes/randomness.html) were explicitly avoided.

TensorFlow determinism evaluation setup

To evaluate the determinism for models built with the TensorFlow library, the same model and training data described for the PyTorch library were used. The TensorFlow experiments were conducted using TensorFlow version 2.2 with multi GPU support provided by the *tf.distribute.MirroredStrategy()* function.

To evaluate determinism on the TensorFlow library, three different setups were employed:

- *Random*: No seeds were set.
- *Seeds*: Random seeds were set for NumPy, TensorFlow, Python's Random module and the Python hash seed.
- *Deterministic*: All random seeds of the setup *seeds* were set together with the *TF_DETERMINISTIC_OPS* environment variable (Supplementary Figure 8).

Moreover, all non-deterministic functions identified by NVIDIA (https://github.com/NVIDIA/framework-determinism) in TensorFlow 2 were avoided.

XGBoost determinism evaluation setup

The forest cover type using the Covertype dataset[53], was used to evaluate determinism for the XGBoost library.XGBoost models were trained with the hyperparameter *subsample* (subsample ratio of the training instances) of 0.5, *colsample_bytree* (subsample ratio of columns) of 0.5 and *colsample_bylevel* (subsample ratio of columns for each level) of 0.5 to introduce random factors.

Additionally, some experiments used the parameter *single_precision*, which reduces the precision when building histograms from double to single. This usually leads to faster convergence at the cost of precision.

Prior to XGBoost 1.1.0 the *hist* histogram building algorithm and its GPU accelerated counterpart *gpu_hist* were susceptible to accumulations of floating-point errors (https://github.com/dmlc/xgboost/issues/5632). Later releases feature deterministic histogram building through floating-point pre-rounding techniques (https://github.com/dmlc/xgboost/pull/5361). XGBoost models were either trained with the officially distributed 1.0.2 version, a CUDA 9 compiled 1.1.0 version or a CUDA 10 compiled 1.1.0 version. Multi GPU support was provided by a *LocalCUDACluster* with Dask (https://dask.org) version 2.14.0, Dask Cuda version 0.13.0 and Dask-ML version 1.4.0.

Hence, the following setups were used to evaluate determinism:

- *No seeds*: Since XGBoost uses a default seed of 0 a randomly picked seed was set for XGBoost.
- Seeds: random seeds were set for Numpy, XGBoost, the XGBoost model's seed, Python's random module and the Python hash seed.
- Single-precision: *single_precision_histogram* was enabled (Supplementary Figure 8).

XGBoost's *allreduce* operations were avoided since they have not yet been verified to run deterministically (https://github.com/dmlc/xgboost/issues/5023) .

*Liver tumor semantic segmentation with PyTorch*

This semantic segmentation analysis was based on the training dataset of the *Liver Tumor Segmentation Benchmark* (LiTS)[40], consisting of 131 abdominal CT scans of patients with HCC, and the corresponding ground-truth segmentation masks of liver and tumor lesions, as annotated by trained radiologists. The tomograms are a mix of pre- and post-therapy abdomen scans, from different CT scanners, using different acquisition protocols, and with a varied number and size of tumors. The segmentation masks are volumetric images where each voxel has a value denoting a class label (0 for background, 1 for liver tissue, and 2 for tumor). The annotated dataset can be downloaded from *codalab* (https://competitions.codalab.org/competitions/17094), or accessed via torrent (https://academictorrents.com/details/27772adef6f563a1ecc0ae19a528b956e6c803ce).

The provided 131 tomograms and segmentation masks with image sizes of 512x512 in the X and Y dimensions, and a variable size in Z (74 - 987), were curated to satisfy memory restrictions. They were first down-scaled by a factor of 2 (to resolutions ranging from 1.12 to 2.0 mm in X and Y, and 0.9 to 12.0 mm on Z) using the *skimage.transform.downscale_local_mean* function of the Scikit-image Python library (https://scikit-image.org/), down-sampling was performed using local averaging. Subsequently, volumes of size 128x128x128 located on the geometric center of the liver annotation, in both the tomograms and segmentation masks, were extracted to create the analyzed dataset.

We used a 3D U-Net architecture[21] (considering 1 input channel and 3 classes), inspired by the reduced model employed by DeepFinder[54]. Accordingly, our model has only two down-sampling stages and contains supplementary convolutional layers in the lowest stage to consider a large spatial context (Supplementary Figure 9). Additionally, we used ReLUs as activation functions and trained the models for 1000 epochs using the ADAM algorithm[55], with a learning rate of 0.0001 and weight decay of 0.0001. We set the batch size to 4 and used dropout with a rate of 0.25. A focal loss[56] was used with an alpha vector (class weights) of (0.2, 1.0, 2.5) and a gamma value of 2.

After being revealed as non-deterministic by the mlf-core linter, we modified our U-Net model to use convolutional layers with a stride of 2 for down-sampling, while the up-sampling operation was performed with the nearest neighbor algorithm. These operations are usually implemented using max-pooling layers and up-convolutions (transposed-convolution), respectively. These modifications were necessary since the used PyTorch library (version 1.7.1) does not offer deterministic implementations of the *torch.nn.MaxPool3d* and *torch.nn.ConvTranspose3d* operations. The model is openly available under https://github.com/mlf-core/liver-ct-segmentation.

To evaluate reproducibility we compared the loss values during training runs (Supplementary Figure 6). We also compared segmentation predictions by measuring performance on a small test set (10% of the available data, 13 randomly selected tomograms), using the Jaccard index or *Intersection over Union* (IoU) as a similarity metric (against expert ground-truth labels), for all training settings. Additionally, we calculated the standard deviation of voxel-wise softmax values, as produced by the models (Supplementary Figure 9). Finally, we calculated the tumor burden for each tomogram in the test set, and measured the standard deviation among experiment runs.

*Analysis of scRNA-seq data with TensorFlow*

Analysis of scRNA-seq data was performed using the Python Scanpy library[57] (v1.6.0) and the *3k PBMCs from a Healthy Donor* dataset available from 10X Genomics (https://support.10xgenomics.com/single-cell-gene-expression/datasets/1.1.0/pbmc3k). The data was downloaded using the Scanpy function *sc.datasets.pbmc3k()*. Cells with less than 200 genes and genes found in less than 3 cells were removed. Counts were normalized to library size (1000 counts per cell) and transformed to logarithmic space. Finally, the count matrix was subsetted to highly variable genes and scaled to zero mean and maximum values of 3 using Scanpy-inbuilt functions.

A simple autoencoder model with layer sizes [256, 128, 64, 32, 64, 128, 256] was fitted to the data. Optimization was performed using the Adam algorithm and a mean squared error loss, a fixed learning rate of 0.001 and a batch size of 256. We fixed the training to 1000 epochs per run.

For comparison of different training settings, we compared the loss per epoch between runs. UMAP embedding was done by extracting the encoding stored in the 32-units layer of the autoencoder and applying the UMAP algorithm[31]. Similarly, for clustering, we extracted the latent space embedding and used the Scanpy functions *sc.pp.neighbors()* and *sc.tl.leiden()* to generate clusters with the Leiden algorithm [58]. To evaluate how differences in the embedding affect cell clustering, we compared the cluster sizes between runs, as cells assigned to different clusters at different runs will lead to cluster size differences. The model is openly available under https://github.com/mlf-core/sc-autoencoder.

*Use case XGBoost - Liver cancer data model*

To train the XGBoost model, gene expression data was collected from 374 hepatocellular carcinoma samples in the Cancer Genome Atlas Liver Hepatocellular Carcinoma (TCGA-LIHC)[59] cohort, including 50 healthy liver samples from the same project; and 136 healthy liver samples of the Genotype-Tissue Expression (GTEx) project[60]. Gene count data of both projects was obtained through the recount2 project portal (https://jhubiostatistics.shinyapps.io/recount/) [61].

The gene counts were transformed into transcript per million (TPM) values using the median transcript length for each gene, which was determined using GTFtools[62].

The genes features used to train the model were reduced to the 556 genes present in the Kyoto Encyclopedia of Genes and Genomes (KEGG) pathway hsa05200 with the title "Pathways in cancer - Homo sapiens (human)"[63]. The data was split into a 75% training and a 25% testing subset.

An XGBoost classification model was fitted to the dataset, and trained for 1000 epochs. The other hyperparameters of the model were *objective: binary:logistic, colsample_bytree: 0.6, learning_rate: 0.2, max_depth: 3, min_child_weight: 1, subsample: 0.7*. Unlisted parameters were left at their default values. The model's hyperparameters were selected following an exhaustive grid search, in which the best hyperparameters were selected based on the highest Matthews correlation coefficient (MCC) value in a 5-fold cross-validation on the training dataset. The model is openly available under https://github.com/mlf-core/lcep.

**Acknowledgements**

We thank Simon Heumos, Tobias Langes and Leon Bichmann for helpful comments on the manuscript. We thank Arjun Jain for fruitful discussions on computer vision. This work was supported by the BMBF-funded de.NBI Cloud within the German Network for Bioinformatics Infrastructure (de.NBI) (031A537B, 031A533A,

031A538A, 031A533B, 031A535A, 031A537C, 031A534A, 031A532B). TensorFlow, the TensorFlow logo and any related marks are trademarks of Google Inc. PyTorch, the PyTorch logo and any related marks are trademarks of Facebook, Inc.  S.N. acknowledges funding by the Sonderforschungsbereich SFB/TR 209 "Liver cancer" of the DFG, from the DFG Project ID 398967434 – TRR 261 and the DFG im Rahmen der Exzellenzstrategie des Bundes und der Länder EXC 2180 – 390900677 and EXC 2124 – 390838134.

**Author contributions**

LH designed and directed the study, conducted the determinism experiments and implemented the mlf-core ecosystem. PE and EM contributed to the mlf-core ecosystem. KM implemented the mlf-core single cell autoencoder use case, LKC conducted determinism experiments, and implemented the use case for liver-tumor segmentation in CT scans, SL implemented the cancer classification from transcriptomics data use case. GG and SN supervised the work. All authors wrote the manuscript, provided critical feedback and helped shape the research, analysis and manuscript.

## Conflict of interest

The authors declare no conflict of interest.

## Data and code availability statement

The complete data together with the implementations of the mlf-core ecosystem and all during this work introduced models are available in the corresponding repositories (see Methods) at [https://github.com/mlf-core](https://github.com/mlf-core).

# Supplementary

*Tables*

| System | Hardware |
| --- | --- |
| 1 - deNBI k80 | Intel 12 core and 2 NVIDIA Tesla K80s |
| 2/3 - denBI V100 | Intel 24 core and 2 NVIDIA V100s |

*Table 1. Overview about used hardware systems for determinism and runtime evaluations.*

*Figures*

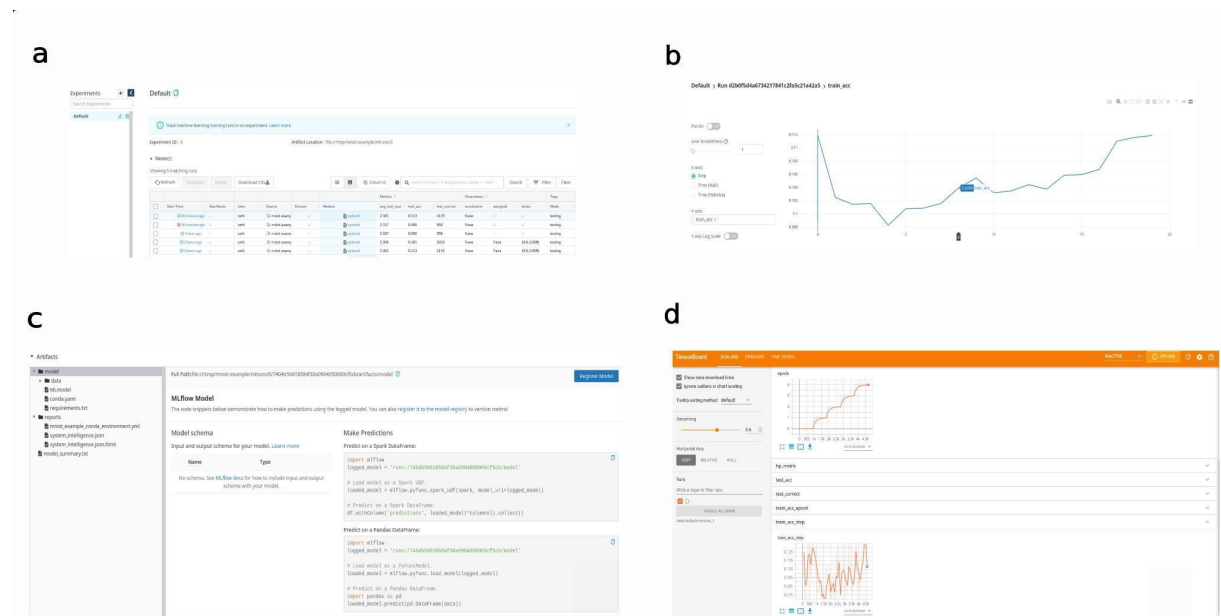

*Supplementary Figure 1. Automatically generated interactive visualizations of the model training process with MLflow (a-c) and Tensorboard (d). (a) Overview of all currently running, failed and successfully completed model training runs associated with all hyperparameters and metrics. (b) Line plot of the training accuracy of an example training run per epoch. (c) Model artifacts of an example run including the pickled model and the system-intelligence hardware report. (d) Training accuracy per step of an example training run.*

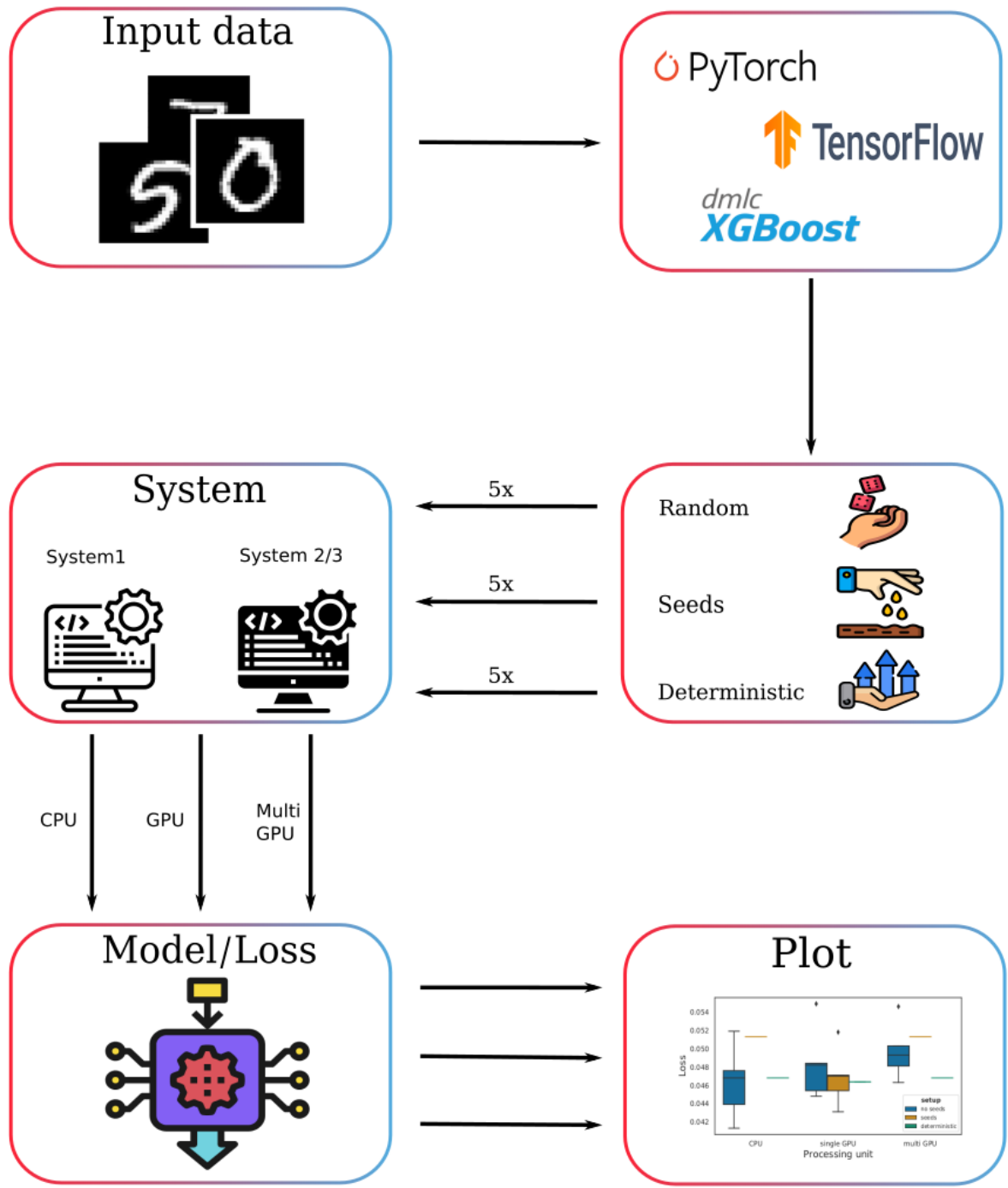

*Supplementary Figure 2. Experimental setup for the determinism evaluation. PyTorch, TensorFlow and XGBoost models were trained with 3 setups. A* random *setup which did not specify any seeds, a* seeds *setup which set all seeds for the respective machine learning library and any underlying libraries and a* deterministic *which additionally forced deterministic algorithms and disabled* cuDNN benchmark. *All setups were trained 5 times for 3 hardware systems. System 2 and 3 shared the same CPU and GPU architectures. This setup was repeated for the CPU, a single GPU and multiple GPUs leading to several models and the corresponding losses.*

Heumos et al.

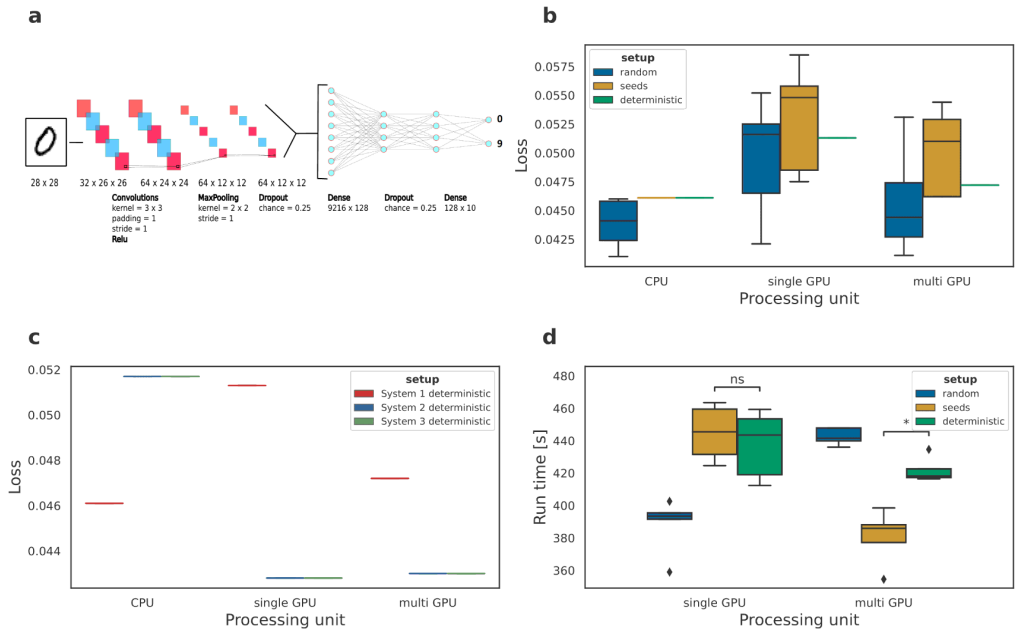

*Supplementary Figure 3. Determinism evaluation of a simple convolutional neural network with dropout layers implemented in PyTorch 1.5 and trained on the MNIST dataset. (a) Neural network architecture for the PyTorch and TensorFlow determinism evaluation. (b) Loss variation across 5 model training runs in the same system with no random seeds or deterministic algorithms (random), solely setting the same library random seeds across runs (seeds); or setting the random seeds and enabling the deterministic algorithms (deterministic). (c) Loss comparison of 5 runs across individual systems with different hardware (systems 1 and 2/3), and individual systems with the same hardware (systems 2 and 3) with deterministic settings. (d) Training time for 25 epochs when training the model without setting random seeds, when setting the random seeds and when enabling the deterministic algorithms.*

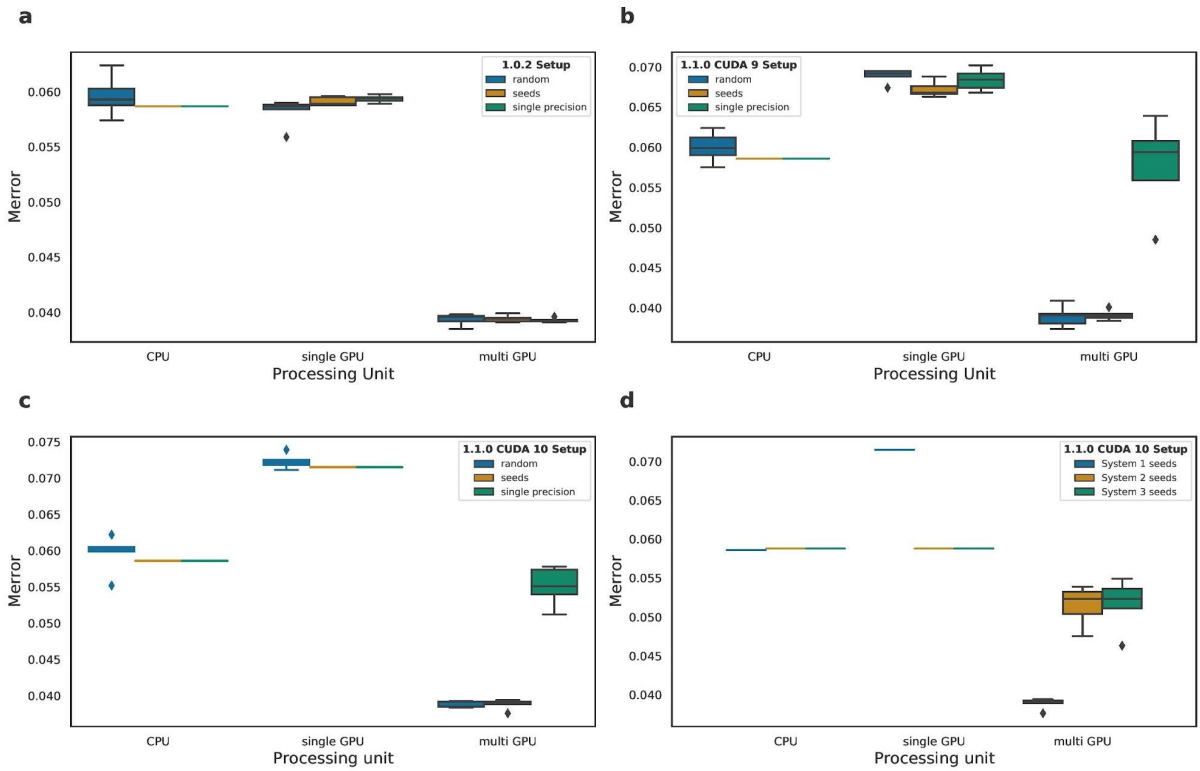

*Supplementary Figure 4. Determinism evaluation of a simple Gradient Boosted Tree implemented in different XGBoost versions and trained on the Covertype dataset. (a) Multiclass classification error rate (Merror) variation across 5 model training runs (N=5) in the same system with no random seeds or deterministic algorithms (random); solely setting the same library random seeds across runs (seeds); or setting the random seeds and enabling single precision for XGBoost with XGBoost version 1.0.2 (single-precision). (b) XGBoost version 1.1.0 compiled with CUDA 9 and (c) XGBoost 1.1.0 compiled with CUDA 10. (d) Merror comparison of 5 runs across individual systems with different hardware (system 1 and 2/3) and individual systems with the same hardware (systems 2 and 3) with all required random seeds set.*

Heumos et al.

a
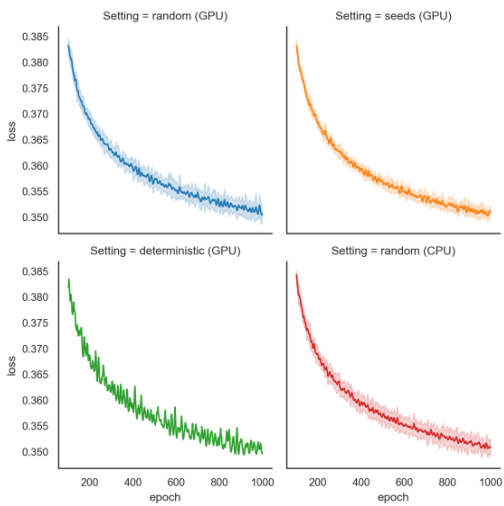

b
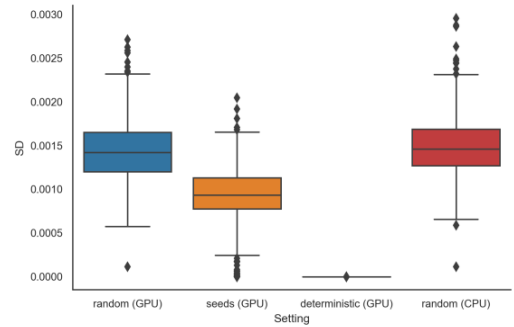

c
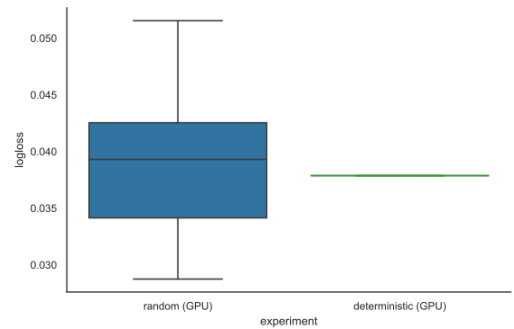

d
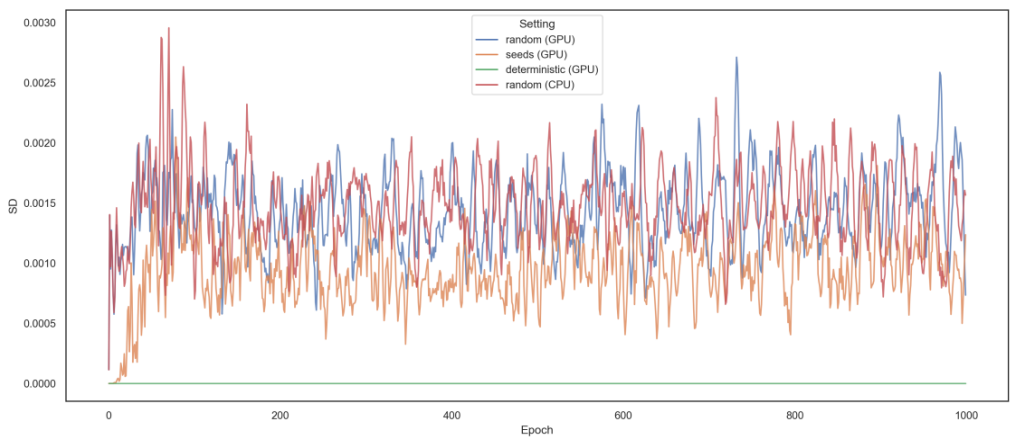

*Supplementary Figure 5. Differences in standard deviations between losses of ten different runs (N=10) in different experimental settings for the single-cell autoencoder and liver cancer feature importance use cases.* (a) *Epoch-wise losses are shown for different experimental settings during training (starting from epoch 100). The standard deviation (SD) between separate runs in one setting are indicated with error bands.* (b) *Boxplots of the SDs of losses at every epoch, for every run and experimental setting.* (c) *Loss (logloss) after 1000 epochs of the XGBoost liver cancer classifying model. For each setup, ten runs were performed.* (d) *SD per epoch plotted for every experimental setting and epoch. The flat green line at the bottom represents the deterministic setting.*

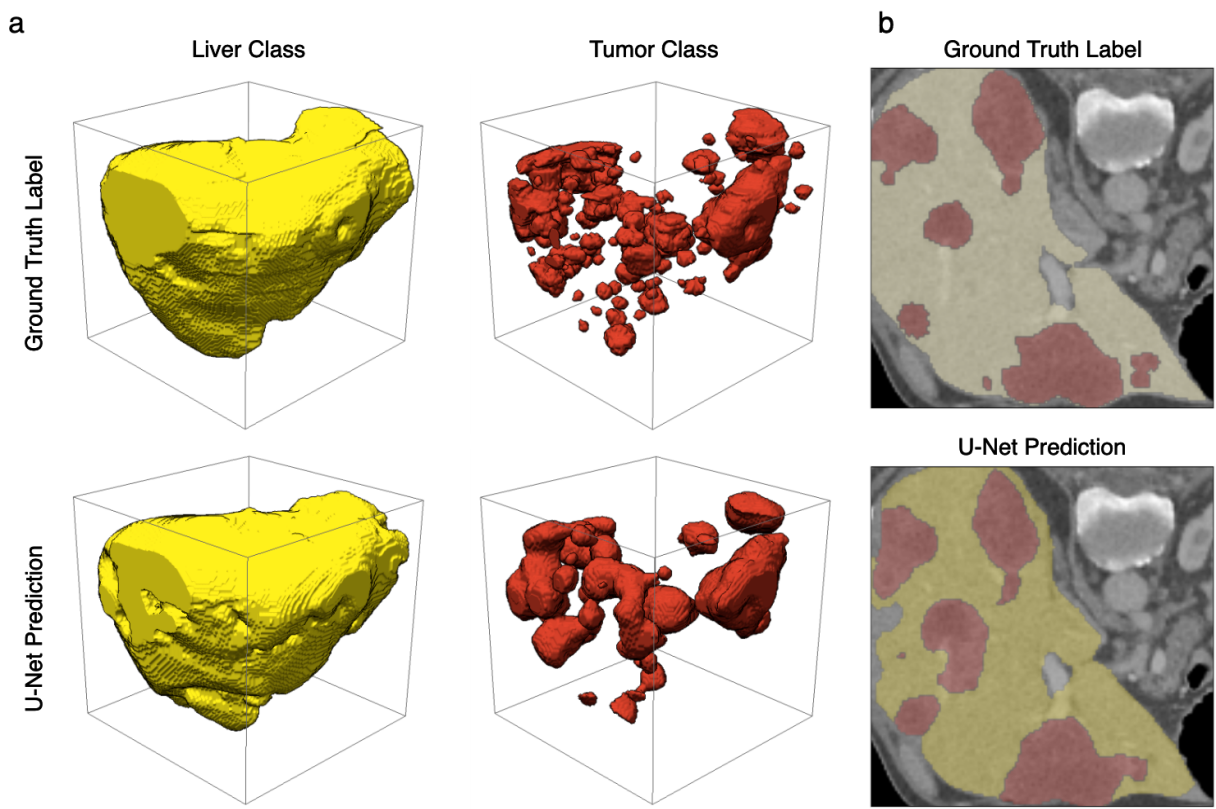

*Supplementary Figure 6. Prediction results of a deterministically trained U-Net model on the LiTS dataset. (a) Volumetric renderings of ground truth labels for the liver class (yellow) and tumor class (red) voxels of a tomogram from the test set, with the corresponding predictions from the deterministically trained model (lower row). (b) Cross-section of the same tomogram, with ground truth labels and model predictions (liver class in yellow, tumor class in red).*

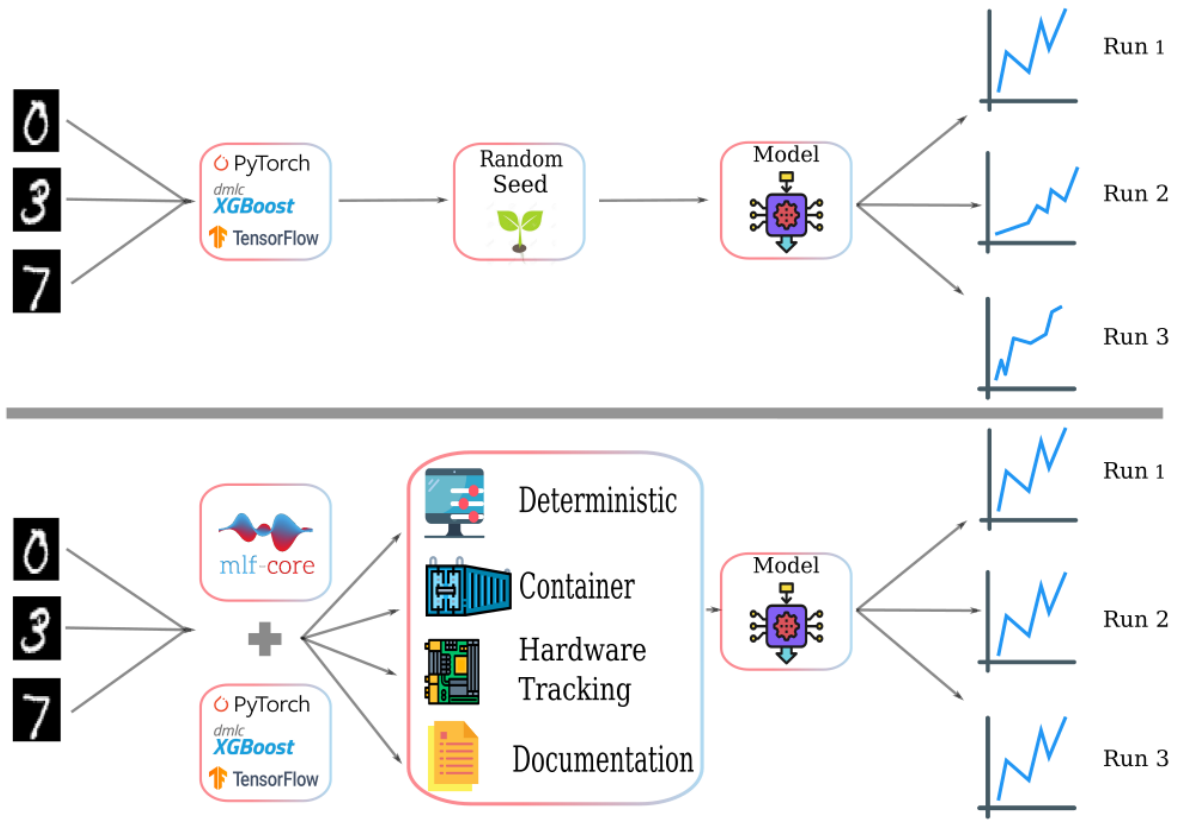

*Supplementary Figure 7. Training models with common machine learning libraries will all random seeds set leads to non-deterministic behavior (top). Contrary, training a model with mlf-core ensures that all required random seeds are set and deterministic algorithms are forced with the mlf-core linter, the runtime environment is containerized by leveraging Conda and Docker containers, the hardware is tracked using system-intelligence and the model is fully documented with a Sphinx documentation setup together with the full training history as recorded with MLflow, leading to fully reproducible machine learning (bottom).*

**a**
```
1  import numpy as np
2  import torch
3  import os
4  import random
5
6  os.environ['PYTHONHASHSEED'] = SEED
7  random.seed(SEED)
8  np.random.seed(SEED)
9  torch.manual_seed(SEED)
10 torch.backends.cudnn.deterministic = True
11 torch.backends.cudnn.benchmark = False
12 # torch.set_deterministic(True)
```

**b**
```
1  import numpy as np
2  import tensorflow as tf
3  import os
4  import random
5
6  os.environ['PYTHONHASHSEED'] = SEED
7  random.seed(SEED)
8  np.random.seed(SEED)
9  tf.random.set_seed(SEED)
10 os.environ['TF_DETERMINISTIC_OPS'] = '1'
11 session_config.intra_op_parallelism_threads = 1
12 session_config.inter_op_parallelism_threads = 1
```

**c**
```
1  import numpy as np
2  import os
3  import random
4
5  os.environ['PYTHONHASHSEED'] = SEED
6  random.seed(SEED)
7  np.random.seed(SEED)
8  param = {'seed': SEED,
9           'single_precision_histogram': True}
```

*Supplementary Figure 8. Set seeds and settings for the PyTorch, TensorFlow and XGBoost determinism evaluation. (a) All explicitly set seeds and PyTorch specific settings.* torch.backends.cudnn.deterministic = True *forces deterministic algorithms whenever available and* torch.backends.cudnn.benchmark = False *disables cuDNN benchmark which is cuDNN's inbuilt auto-tuner to find the fastest algorithms for the used hardware. Line 12 was not used for the evaluation since it was not available yet, but is used for the most recent mlf-core PyTorch project template.*
*(b) All explicitly set seeds and TensorFlow specific settings.* torch.backends.cudnn.deterministic = True *forces deterministic algorithms when available and* torch.backends.cudnn.benchmark = False *disables cuDNN's inbuilt auto-tuner to find the fastest algorithms for the used hardware. (c) All explicitly set seeds and XGBoost specific settings.* single_precision_histogram *enables single precision histogram building to speed up training.*

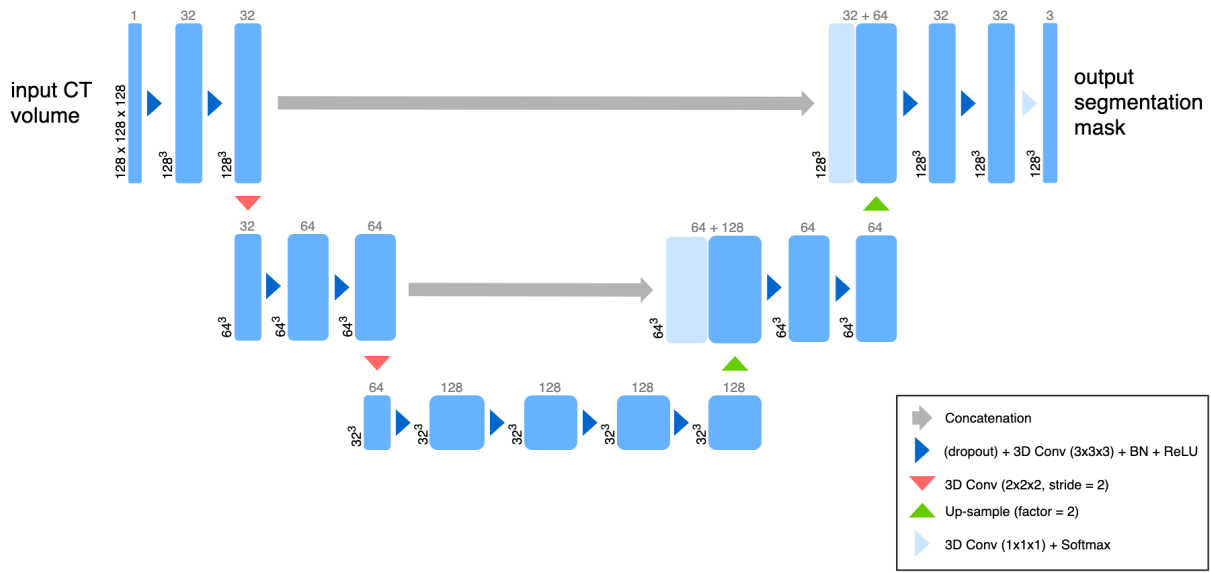

*Supplementary Figure 9. A reduced 3D U-Net architecture. The U-Net is a convolutional "encoder-decoder" model for semantic segmentation of 2D and 3D images[21,39]. In this simplified model, convolutional layers with a stride of 2 are used for down-sampling, while the up-sampling operation was performed with the nearest neighbor algorithm. Here, convolutions use filter sizes of 3x3x3, dropout is applied to every convolutional layer, and the softmax function is used on the last layer to produce class pseudo-probabilities. Blue boxes correspond to 3D multi-channel feature maps, with the number of channels denoted on top, and the size of the spatial dimensions marked in the lower left.*

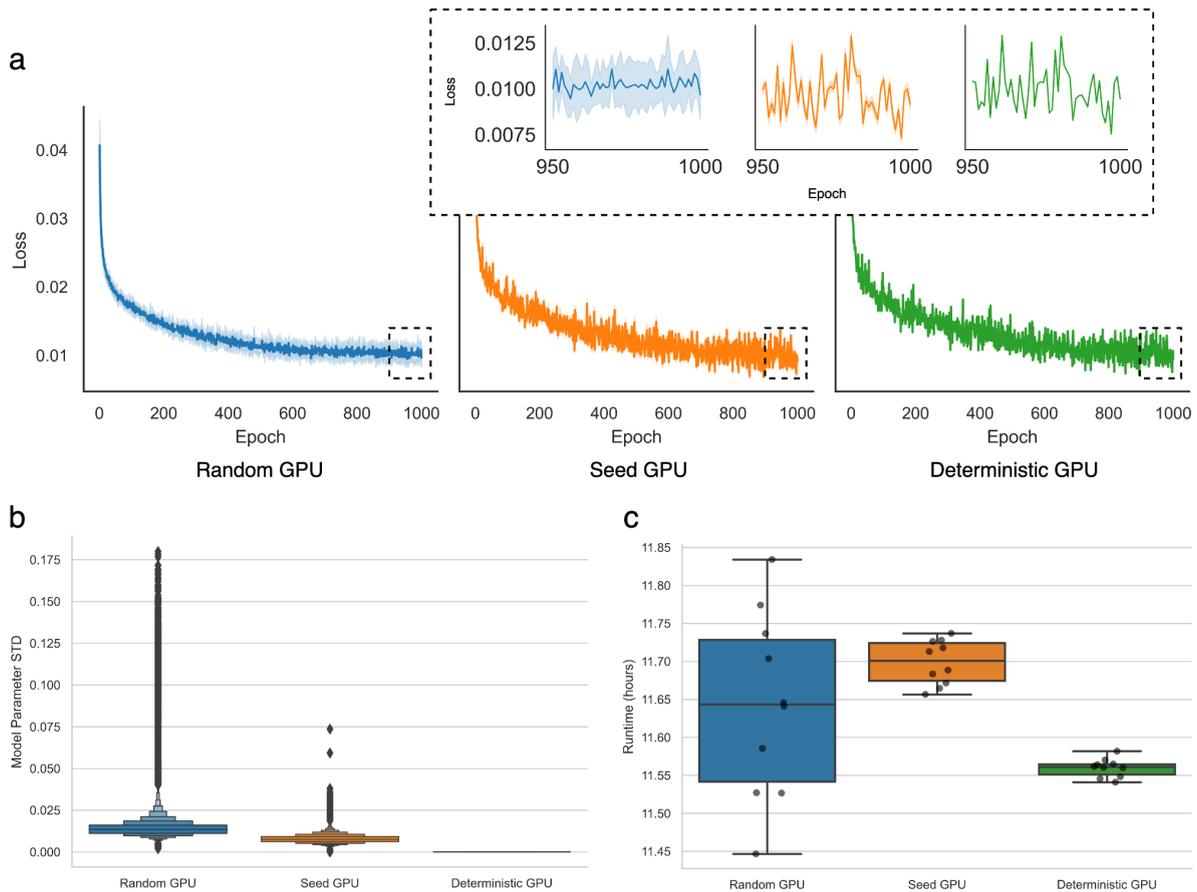

*Supplementary Figure 10. Epoch-wise differences in loss between training runs (N=10) in different experimental settings (Random, Seed, Deterministic) for the U-Net model trained on the LiTS dataset, and deviation of model parameters for each of the settings. (a) Loss value during training is shown for different experimental settings, the standard deviation between separate runs is indicated with error bands. The dashed lines mark the zoomed in regions of the curves, corresponding to the last 50 epochs. (b) Letter-value plot of standard deviation values of U-Net model parameters (weights and biases) across training runs (resulting models), the standard deviation of all 3,682,595 parameters was calculated for all experimental settings. (c) Training runtimes (1000 epochs on a model with 3,682,595 trainable parameters) for runs of all experimental settings. All training runs were computed using 2 GPUs, each an NVidia Tesla V100 (32 GB).*

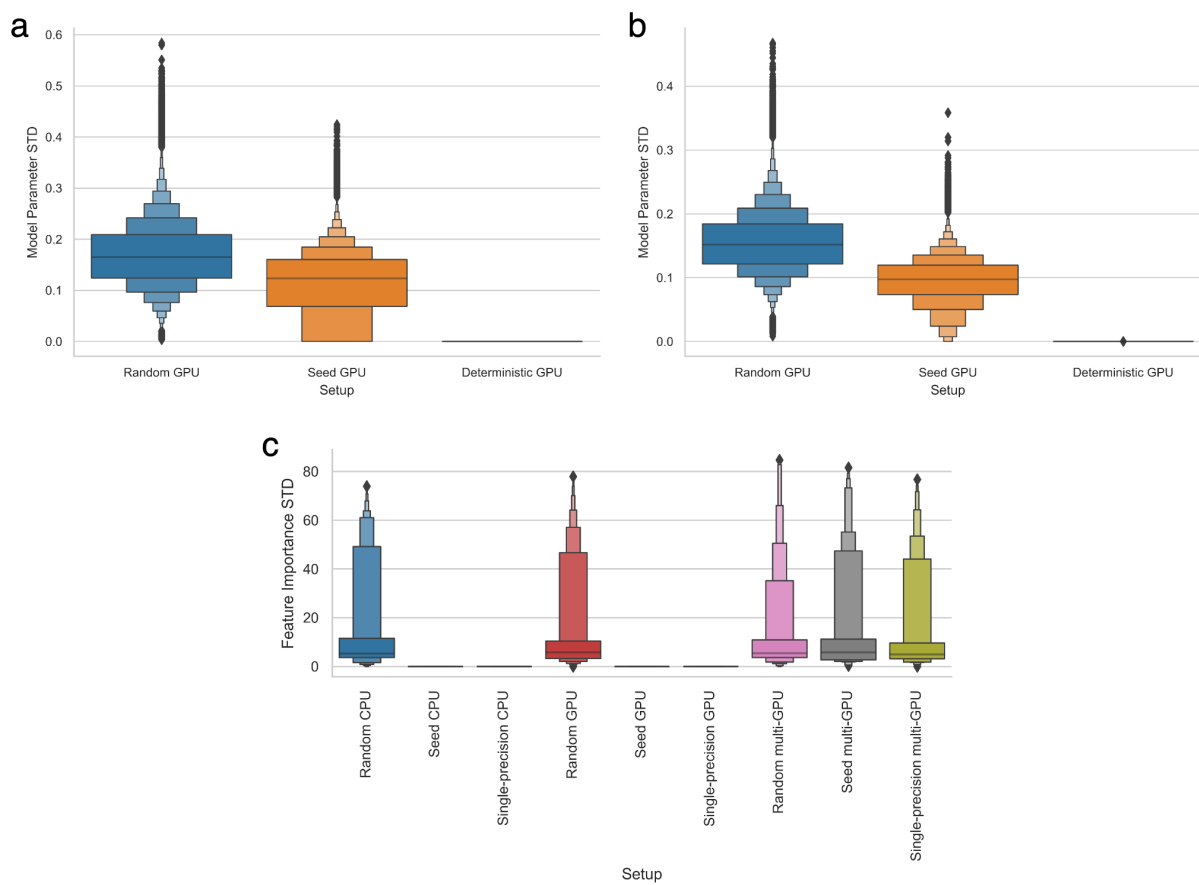

*Supplementary Figure 11. Deviation of model parameters (weights and biases) and feature importance values after training runs, for the MNIST (PyTorch, TensorFlow) and CovType (XGBoost) datasets, in different experimental settings (setup). Models were trained 10 times on each setup (N=10). (a) Letter-value plot of standard deviations values of model parameters, for PyTorch models trained (100 epochs) on the MNIST dataset. (b) Letter-value plot of standard deviations values of model parameters, for Tensorflow models trained (100 epochs) on the MNIST dataset. (c) Letter-value plot of standard deviations of feature importance values from XGBoost models, trained on the CovType dataset.*